\documentclass[draft]{agujournal2019}
\usepackage{url} 
\usepackage{lineno}
\usepackage[inline]{trackchanges} 
\usepackage{soul}
\usepackage{multirow}
\usepackage{amsmath,amssymb}

\draftfalse

\journalname{Geophysical Research Letters}

\begin{document}

\title{Ion-Acoustic Waves and the Proton-Alpha Streaming Instability at Collisionless Shocks}

\authors{D. B. Graham \affil{1}, Yu. V. Khotyaintsev\affil{1}, A. Lalti\affil{2}}

\affiliation{1}{Swedish Institute of Space Physics, Uppsala, Sweden.}
\affiliation{2}{Northumbria University, Newcastle upon Tyne, UK.}
\correspondingauthor{D. B. Graham}{dgraham@irfu.se}

\begin{keypoints}
\item The proton-alpha streaming instability caused by the cross-shock potential can excite ion-acoustic waves.
\item Theory and a simulation of the instability yield results consistent with observations. 
\item The instability results in ion heating and reduces the relative drift between ion species.
\end{keypoints}

\begin{abstract}
Ion-acoustic waves are routinely observed at collisionless shocks and could be an important source of resistivity. The source of instability and the effects of the waves are not fully understood. We show, using Magnetospheric Multiscale (MMS) mission observations and numerical modeling, that across low Mach number shocks a large relative drift between protons and alpha particles develops, which can be unstable to the proton-alpha streaming instability. The results from linear analysis and a numerical simulation show that the resulting waves agree with the observed wave properties. The generated ion-acoustic waves are predicted to become nonlinear and form ion holes, maintained by trapped protons and alphas. The instability reduces the relative drift between protons and alphas, and heats the ions, thus providing a source of resistivity at shocks. 
\end{abstract}

\section*{Plain Language Summary}
Shock waves in collisionless plasmas are a fundamental process, which can produce intense particle acceleration. These shocks require a source of resistivity, which can be produced by plasma waves, to be sustained. A wide variety of waves have been reported at collisionless shocks, although the sources of the waves and their effects on the plasma are still debated. Using observations from the Magnetospheric Multiscale (MMS) spacecraft and numerical modeling, we show that a large drift between protons and alpha particles occurs across the shock, and leads to the generation of large-amplitude electrostatic waves. These waves subsequently heat the protons and alpha particles, and reduce the relative drift between the species. The waves can provide an important source of the resistivity required to sustain shocks. 


\section{Introduction}
Shock waves are ubiquitous in solar and astrophysical plasmas. Shocks in plasma are often collisionless, meaning particle collisions cannot provide the resistivity required to sustain them. Instead, other sources of resistivity are required, such as wave-particle interactions \cite{sagdeev1966,papadopoulos1985} and ion reflection \cite{edminston1984}. Observations and simulations have shown that a wide variety of large-amplitude electrostatic and electromagnetic waves develop across shocks \cite{wu1984}. 

Large-amplitude electrostatic ion-acoustic waves are routinely observed at collisionless shocks \cite{boldu2024,wilson2007}. At Earth's bow shock, ion-acoustic waves account for the largest electric fields \cite{fredricks1970,bale2007}, reaching up to $\sim 1$~V~m$^{-1}$ \cite{vasko2022}. These waves have been observed at both high \cite{goodrich2019,vasko2022} and low Mach number shocks \cite{boldu2024,wilson2007}. Observations have shown that the waves are often highly oblique to the background magnetic field \cite{hull2006}. Additionally, nonlinear solitary structures have been found, consisting of ion \cite{hobara2008,vasko2020,wang2020b,wang2021b} and electron phase-space holes \cite{bale1998,kamaletdinov2022}. These waves and nonlinear structures have been argued to provide an important contribution to anomalous resistivity \cite{wilson2014} and particle heating \cite{kamaletdinov2024,vasko2020}. 

Various instabilities have been proposed for the generation of ion-acoustic waves at shocks, including the proton-proton streaming instability \cite{akimoto1985,formisano1982,gary1987a,goodrich2019}, Buneman instability \cite{buneman1959}, and instabilities associated with temperature gradients \cite{allan1974}. 
The observation of ion holes rather than electron holes in some bow shock crossings suggests that proton-proton streaming instabilities are responsible for the waves, rather than the Buneman instability or electron-electron streaming instabilities \cite{vasko2020}. However, instability from alpha particles has not been considered in detail. 

We investigate the generation of ion-acoustic waves at five low Mach number quasi-perpendicular bow shock crossings observed on 24 April 2023 and argue that they are generated by the proton-alpha streaming instability. The outline of this letter is: In section \ref{observationssec} MMS observations are presented, in section \ref{shockmodel} we model the proton and alpha distributions across the shocks, in section \ref{seclintheory} we calculate the instability properties using linear theory, and in section \ref{secpicmodel} we model the nonlinear evolution of the instability. 
Sections \ref{secdiscussion} and \ref{secconclusions} are the discussion and conclusions. 




\section{Observations and Overview} \label{observationssec}
We use data from the MMS spacecraft \cite{burch2016a}: Magnetic field ${\bf B}$ from the Fluxgate Magnetometer (FGM) \cite{russell2016}, electric field ${\bf E}$ and probe potentials from Electric field Double Probes (EDP) \cite{ergun2016,lindqvist2016}, and particle distributions and moments from Fast Plasma Investigation (FPI) \cite{pollock2016}. We use burst mode data from MMS1. 

\begin{figure*}[htbp!]
\begin{center}
\includegraphics[width=160mm, height=96mm]{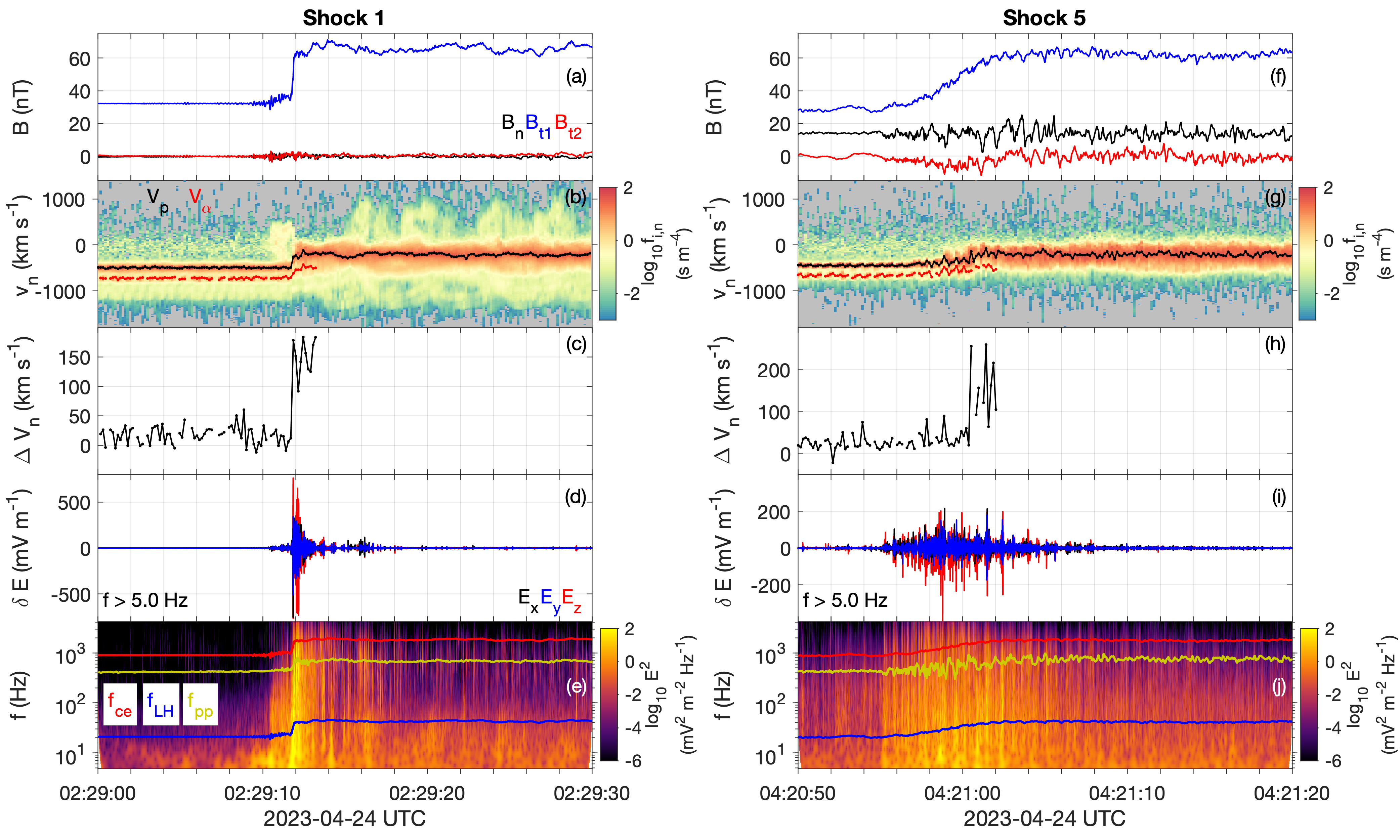}
\caption{Overview of the crossings of shocks 1 and 5 observed by MMS1. (a)--(e) Shock 1 and (f)--(j) Shock 5.(a) ${\bf B}$ in $(\hat{\bf n},\hat{\bf t}_1,\hat{\bf t}_2)$ coordinates. (b) Reduced 1D ion distribution along $v_n$. (c) $\Delta V_n = (V_p - V_{\alpha})_n$. (d) $\delta {\bf E}$ in despun local (DSL) coordinates. (e) Frequency-time spectrogram of ${\bf E}$. The red, yellow, and blue lines are the electron cyclotron frequency $f_{ce}$, proton plasma frequency $f_{pp}$, and the lower hybrid frequency $f_{LH}$, respectively. Panels (f)--(j) show the same quantities as (a)--(e). }
\label{Figobs1}
\end{center}
\end{figure*}

We investigate the five bow shock crossings observed on 24 April 2023. The shocks were observed as a magnetic cloud proceeding a coronal mass ejection crossing Earth. The shocks are quasi-perpendicular and characterized by low Mach numbers and low plasma beta $\beta$ \cite{graham2024b}. Figure \ref{Figobs1} provides an overview of the first and fifth shock crossings, shocks 1 and 5, respectively. Shock 1 is a subcritical shock characterized by $\theta_{Bn} = 89^{\circ}$ and $M_f = 1.8$ \cite{graham2024b}. Figure \ref{Figobs1}a shows ${\bf B}$ in shock-normal coordinates $(\hat{\bf n},\hat{\bf t}_1,\hat{\bf t}_2)$ calculated using the mixed-mode methods \cite{abraham-shrauner1972}, where $\hat{\bf n}$ is normal to the shock, $\hat{\bf t}_1$ is aligned with the upstream ${\bf B}$, and $\hat{\bf t}_2$ completes the right-hand coordinate system. We observe a small foot in $B_{t1}$ just ahead of the narrow ramp. 

Figure \ref{Figobs1}b shows the one-dimensional (1D) reduced ion distributions, $f_i(v_n)$, along $v_n$. We observe a small fraction of reflected protons in the foot region. FPI measures all ions and sorts them by energy-per-charge, which is converted to speed assuming the proton mass and charge for the reduced distributions. This results in alpha particle speeds being overestimated by $\sqrt{2}$ in the spacecraft frame \cite{graham2024a},  enabling different ions moving at the same speed to be distinguished. Upstream of the shock, we observe protons, alphas, and singly charged Helium ions. For protons and alphas, we estimate their bulk velocity along $\hat{\bf n}$, $V_n$, in the spacecraft frame from the peaks in $f_i(v_n)$ in Figure \ref{Figobs1}b (black and red lines). We distinguish protons and alphas upstream of the shock and just behind the ramp. It is difficult to distinguish protons and alphas further downstream of the ramp. From these observations, we calculate $\Delta V_n = V_{p,n} - V_{\alpha,n}$ (Figure \ref{Figobs1}c). Upstream of the shock $\Delta V_n$ is slightly positive, corresponding to alphas moving Earthward slightly faster than protons. Across the ramp $V_{p,n}$ and $V_{\alpha,n}$ are decelerated by the cross-shock potential $\phi$, with protons decelerated more than alphas due to their lower mass-to-charge ratio. This results in $\Delta V_n \approx 150$~km~s$^{-1}$ just downstream of the ramp. 

Figures \ref{Figobs1}d and \ref{Figobs1}e show the waveform of the fluctuating electric field $\delta {\bf E}$ and the associated frequency-time spectrogram. At the ramp, we observe electrostatic waves with amplitude reaching $\sim 800$~mV~m$^{-1}$. Figure \ref{Figobs1}e shows that the waves are broadband, with power around the proton plasma frequency $f_{pp}$. 

Figures \ref{Figobs1}f--\ref{Figobs1}j provide an overview of shock 5, characterized by $\theta_{Bn} = 66^{\circ}$ and $M_f = 1.7$. The ramp width is substantially broader than shock 1. Figures \ref{Figobs1}i and \ref{Figobs1}j show large-amplitude $\delta {\bf E}$ and broadband power spectra across the ramp. Figure \ref{Figobs1}g shows that there is no proton reflection for this shock  and no foot is observed. However, Figures \ref{Figobs1}g and \ref{Figobs1}h show that $\Delta V_n$ increases across the ramp and peaks around $200$~km~s$^{-1}$.  

\begin{figure*}[htbp!]
\begin{center}
\includegraphics[width=140mm, height=160mm]{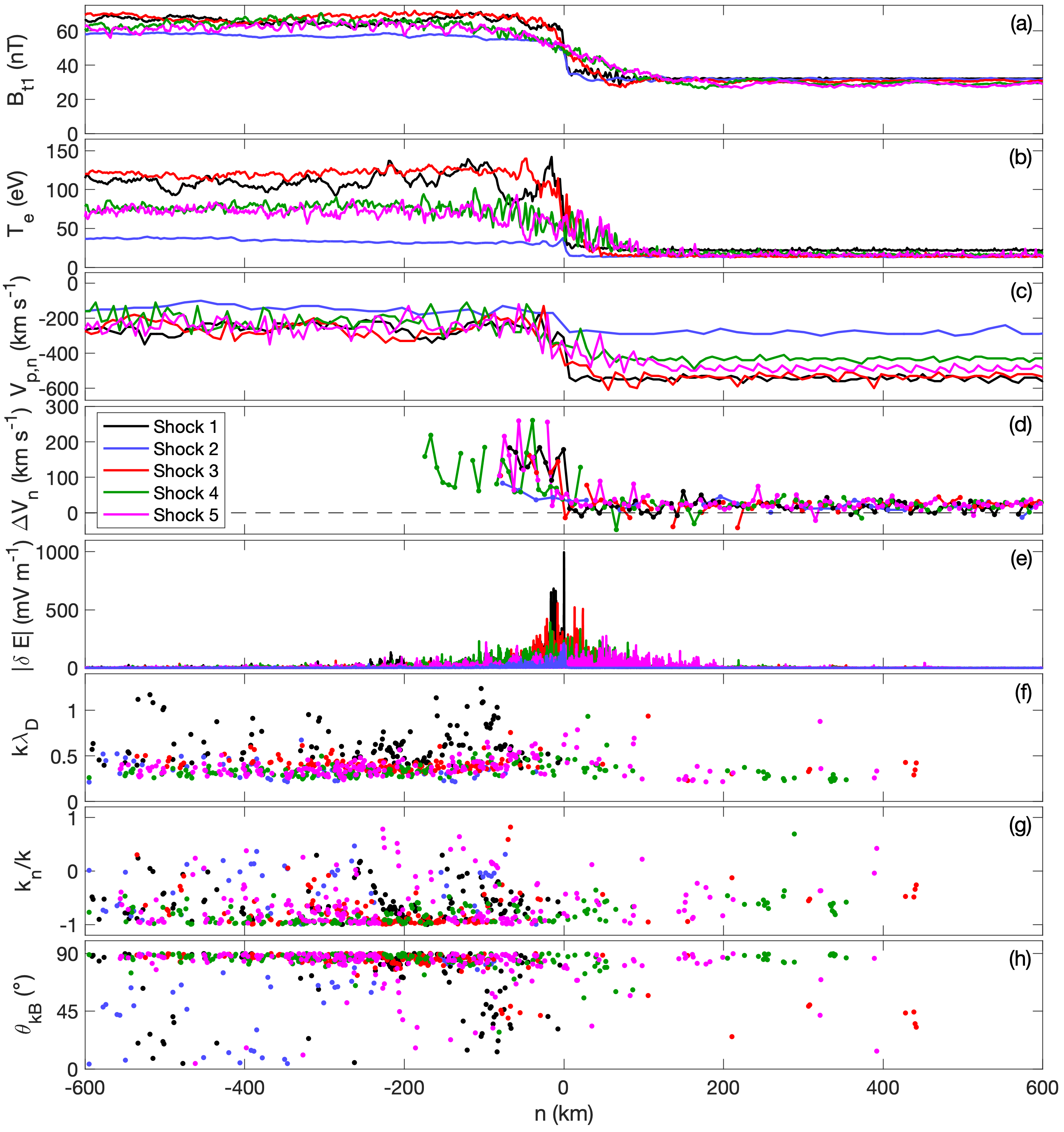}
\caption{Profiles of the five shocks and properties of the ion-acoustic waves, where time is converted to position along $\hat{\bf n}$. (a) $B_{t1}$. (b) $T_e$. (c) $V_{p,n}$. (d) $\delta V_n$. (e) $|\delta E|$ for $f > 200$~Hz. (f)--(h) $k\lambda_D$, $k_n/k$, and $\theta_{kB}$ of ion-acoustic waves. The black, blue, red, green, and magenta curves are shocks 1--5.}
\label{Figfiveshocks}
\end{center}
\end{figure*}

Figure \ref{Figfiveshocks} shows the profiles of the five shocks, and properties of the ion-acoustic waves, near the ramp versus position along $\hat{\bf n}$, estimated from the shock speed \cite{graham2024b}. Figure \ref{Figfiveshocks}a shows $B_{t1}$ of the five shocks. The shock width $l$ increases with shock number. Figures \ref{Figfiveshocks}b and \ref{Figfiveshocks}c show that across the ramps, $T_e$ increases significantly and $V_{p,n}$ decreases. Figure \ref{Figfiveshocks}d shows that upstream of the shocks $\Delta V_n > 0$, but is small. Across the ramp and just downstream $\Delta V_n$ reaches $\sim 200$~km~s$^{-1}$, except for shock 2, where $\Delta V_n \sim 100$~km~s$^{-1}$. Further downstream $\Delta V_n$ can persist, but is difficult to calculate from $f_i(v_n)$. Figure \ref{Figfiveshocks}e shows that for all shocks the largest amplitude $|\delta {\bf E}|$ occurs across the ramp where large $\Delta V_n$ develops. 

Figures~\ref{Figfiveshocks}f--\ref{Figfiveshocks}h show the wave properties calculated using the multi-probe interferometry method in \citeA{lalti2023}. The points correspond to intervals where the dispersion relation and wave vector ${\bf k}$ can be reliably estimated. Similar wave properties are observed across the five shocks. Close to the ramp, the waves are characterized by $k\lambda_D \sim 0.4$ (Figure~\ref{Figfiveshocks}f). Figure~\ref{Figfiveshocks}g shows that ${\bf k}$ is typically closely aligned with $\hat{\bf n}$. Therefore, the waves are highly oblique to ${\bf B}$ (Figure~\ref{Figfiveshocks}g), with most of the waves having wave-normal angles $\theta_{kB} \sim 90^{\circ}$ (Figure~\ref{Figfiveshocks}h). In the plasma frame, the waves are typically characterized by $f/f_{pp} \lesssim 1$ and phase speed $v_{ph} \lesssim c_s$, where $c_s$ is the ion-acoustic speed (not shown). Based on these properties, we conclude that the observed $\delta {\bf E}$ are consistent with obliquely propagating ion-acoustic-like waves and nonlinear electrostatic structures convected by ${\bf V}$.   

\section{Modeling of Ion Distributions} \label{shockmodel}
\begin{figure*}[htbp!]
\begin{center}
\includegraphics[width=160mm, height=96mm]{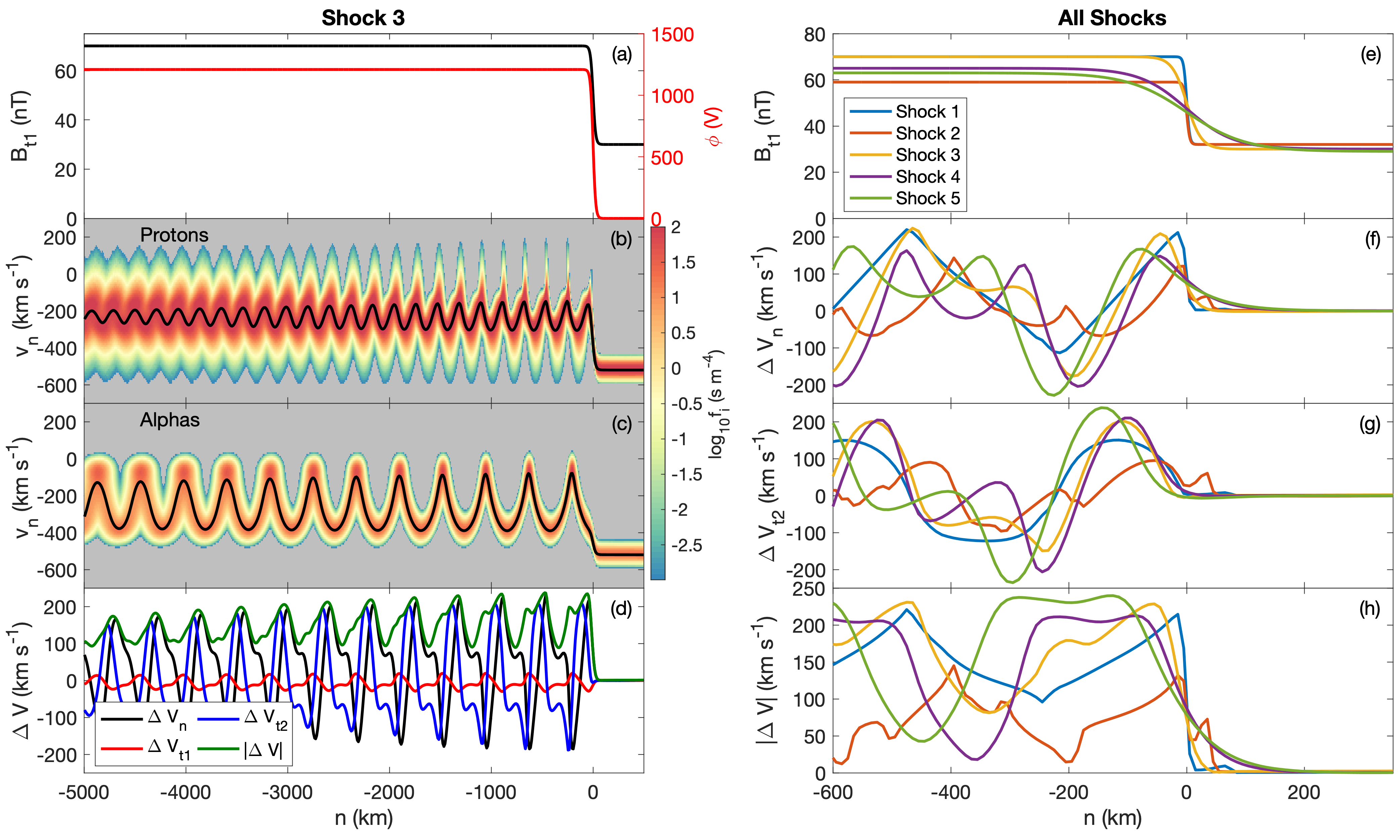}
\caption{Modeled proton and alpha distributions predicted by Liouville mapping. (a)--(d) Predicted bulk velocities and ion distributions for shock 3, (e)--(h) ${\bf V}_p - {\bf V}_\alpha$ for the five shocks. (a) Modeled $B_{t1}$ and $\phi$ for shock 3. (b) and (c) Reduced proton and alpha distributions along $v_n$. The black lines are $V_{p,n}$ and $V_{\alpha,n}$. (d) $\Delta {\bf V}$ in $(\hat{\bf n},\hat{\bf t}_1,\hat{\bf t}_2)$ coordinates and $|\Delta {\bf V}|$. (e) Profiles of $B_{t1}$ for the five shocks. (f)--(h) $\Delta V_n$, $\Delta V_{t2}$, and $|\Delta {\bf V}|$, respectively. For shock 2 we reduced $\phi$ by $10 \, \%$ compared with model prediction in \citeA{graham2024b} to reduce the influence of reflected protons on the velocity moments. }
\label{Lmappingfig}
\end{center}
\end{figure*}

To investigate the difference in bulk velocities between protons and alphas, we use the results of the Liouville mapping model and shock parameters from \citeA{graham2024b}. Figures \ref{Lmappingfig}a--\ref{Lmappingfig}d show the modeled shock profile and predicted proton and alpha distributions for shock 3. In the normal-incidence frame (NIF) the cross-shock potential is $\phi \approx 1200$~V, which reflects a small fraction of protons. Figures \ref{Lmappingfig}b and \ref{Lmappingfig}c show $f_i(v_n)$ for protons and alphas. At the ramp, the ions are rapidly decelerated by $\phi$, with protons decelerated to significantly lower speeds. Downstream of the ramp, the protons and alphas undergo periodic gyrations along $v_n$ and $v_{t2}$, and undergo gradual gyrophase mixing \cite{balikhin2008,gedalin2015}. Due to the differential deceleration across the ramp and distinct gyro-scales of protons and alphas, a large velocity difference $\Delta {\bf V}$ persists far downstream of the shock (Figure~\ref{Lmappingfig}d). Downstream of the ramp, $\Delta {\bf V}$ fluctuates between the $\hat{\bf n}$ and $\hat{\bf t}_2$ directions, and gradually decreases due to gyrophase mixing. 

For very narrow ramp widths, the change in $V_n$ of ions transmitted across the shock ramp without reflection is given by
\begin{linenomath}
\begin{equation}
V_{u,n}^2 - V_{d,n}^2 = \frac{2 Z e \phi}{m_i} \rightarrow V_{d,n} = \sqrt{V_{u,n}^2 - \frac{2 Z e \phi}{m_i}},
\label{dVneq}
\end{equation}
\end{linenomath}
where $Z$ is the ion charge number. Alpha particles are decelerated to a lesser degree than protons. From equation (\ref{dVneq}), the maximum $|\Delta {\bf V}|_n$ is $V_{u,n}/\sqrt{2}$ for $e \phi = m_p V_{u,n}^2/2$. Thus, for large $\phi$ significant $\Delta {\bf V}_n$ develops across the ramp. 

Figures \ref{Lmappingfig}f--\ref{Lmappingfig}h show $\Delta {\bf V}_n$ predicted for the five shocks \cite{graham2024b}. For all shocks, $\Delta {\bf V}_n$ develops across the ramps and peaks at $\Delta {\bf V}_n \approx 200$~km~s$^{-1}$, except for shock 2, which peaks at $\Delta {\bf V}_n \approx 100$~km~s$^{-1}$. These values agree with the observed values in Figure~\ref{Figfiveshocks}d. Downstream of the shocks, $\Delta {\bf V}$ fluctuates in the $\hat{\bf n}$ and $\hat{\bf t}_2$ directions due to the gyromotion of the proton and alpha distributions and persists far downstream. However, if wave-particle interactions strongly affect the ions, the model will be inaccurate downstream of the ramp. 

\section{Linear analysis} \label{seclintheory}
We now investigate the conditions for which $\Delta {\bf V}$ are unstable and generate ion-acoustic waves. We consider the unmagnetized homogeneous dispersion equation \cite{gary1987a}: 
\begin{linenomath}
\begin{equation}
0 = 1 - \frac{\omega_{pp}^2}{k^2 v_p^2} Z' \left( \frac{\omega}{k v_p} \right) - \frac{\omega_{p\alpha}^2}{k^2 v_{\alpha}^2} Z' \left( \frac{\omega - k_{n} V_{\alpha}}{k v_\alpha} \right) - \frac{\omega_{pe}^2}{k^2 v_e^2} Z' \left( \frac{\omega}{k v_e} \right), 
\label{dispeq}
\end{equation}
\end{linenomath}
where $\omega_{ps}$ is the plasma frequency of species $s$, $v_s = \sqrt{2 k_B T_s/m_s}$ is the thermal speed, $m_s$ is the particle mass, $T_s$ is the temperature, $k$ is the wave number, $k_{n}$ is the wave number along the direction of the drift in alpha particles, $V_\alpha$ is the bulk speed of the alpha particles, and $Z'$ is the derivative of the plasma dispersion function \cite{fried1961}. Here $s = p, \alpha, e$ refers to the proton, alpha, and electron distributions. We assume the proton and electron distributions are stationary, while the alpha distribution has a bulk velocity ${\bf V}_{\alpha}$ along the $\hat{\bf n}$-direction. 

\begin{figure*}[htbp!]
\begin{center}
\includegraphics[width=160mm, height=170mm]{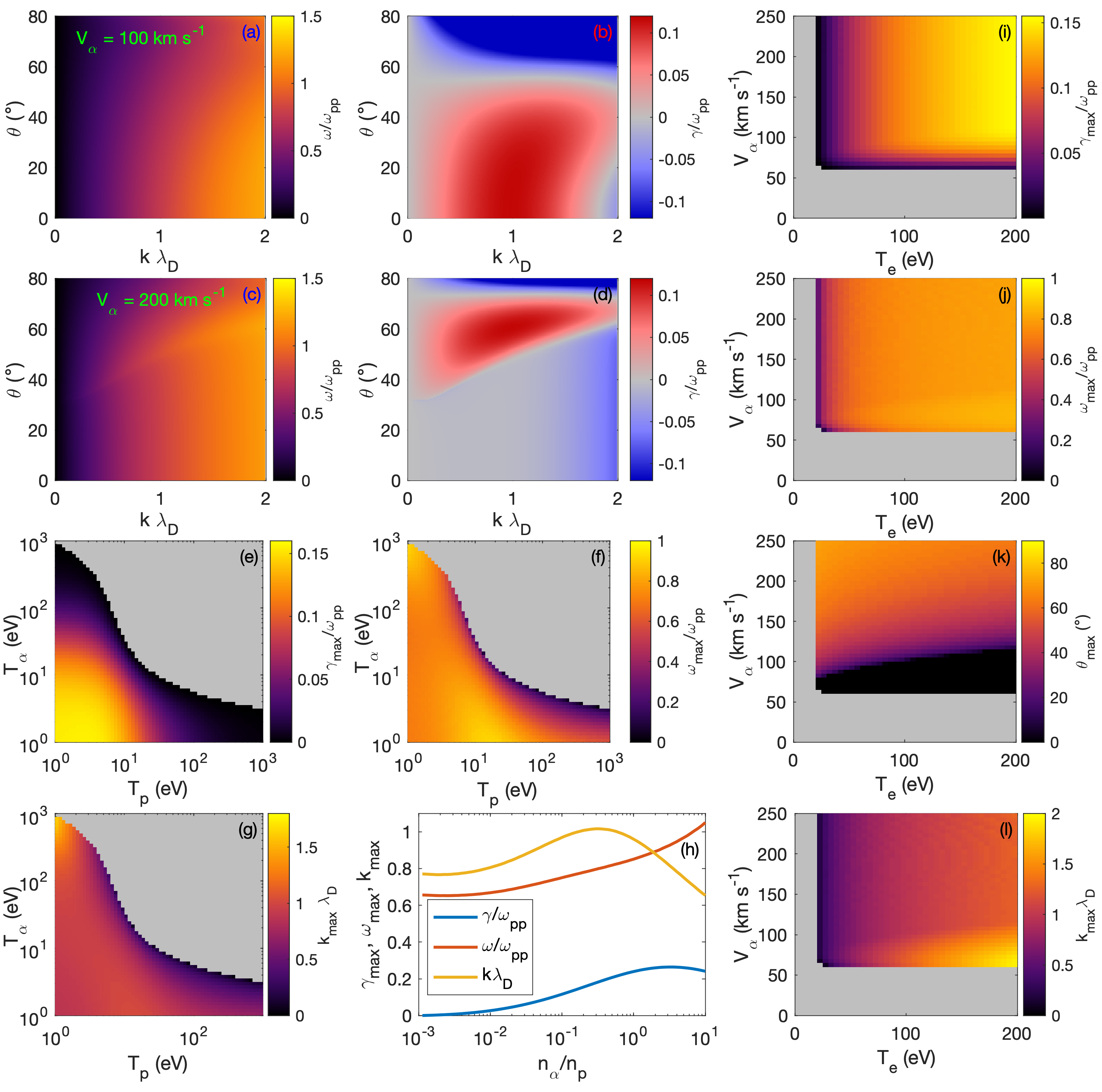}
\caption{Properties of the proton-alpha streaming instability. (a) and (b) Frequency $\omega$ and growth rates $\gamma$ versus $k$ and $\theta$ for nominal conditions. (c) and (d) Same as (a) and (b) for $V_\alpha = 200$~km~s$^{-1}$. (e)--(g) Peak growth rate $\gamma_{\mathrm{max}}$, and $\omega$ and $k$ corresponding to $\gamma_{\mathrm{max}}$, $\omega_{\mathrm{max}}$ and $k_{\mathrm{max}}$ as functions of $T_\alpha$ and $T_p$ for otherwise nominal conditions. (h) $\gamma_{\mathrm{max}}$, $\omega_{\mathrm{max}}$ and $k_{\mathrm{max}}$ as a function of $n_\alpha/n_p$. In panels (e)--(h) $\theta = 0^{\circ}$ is assumed. (i)--(l) $\gamma_{\mathrm{max}}$, $\omega_{\mathrm{max}}$, angle $\theta_{\mathrm{max}}$ corresponding to $\gamma_{\mathrm{max}}$, and $k_{\mathrm{max}}$ as functions of $T_e$ and $V_{\alpha}$. Gray-shaded regions in panels (e)--(l) correspond to no instability.}
\label{Disprelfig}
\end{center}
\end{figure*}

As nominal plasma conditions we use $n_e = 12$~cm$^{-3}$ and $T_e = 100$~eV for electrons, $n_p = 10$~cm$^{-3}$ and $T_p = 3$~eV for protons, and $n_\alpha = 1$~cm$^{-3}$, $T_\alpha = 12$~eV, and $V_\alpha = 100$~km~s$^{-1}$ for alphas. We consider how the instability changes with  temperatures, $n_\alpha/n_p$, and $V_\alpha$. 
Figures \ref{Disprelfig}a and \ref{Disprelfig}b show the real and imaginary frequencies, $\omega$ and $\gamma$, as functions of wave number $k$ and the angle $\theta$ between the wave vector ${\bf k}$ and ${\bf V}_\alpha$ for nominal conditions. Here, $\omega$ depends weakly on $\theta$, and that $\gamma > 0$ for $\theta < 56^{\circ}$. The maximum growth rate $\gamma_{\mathrm{max}} = 0.12 \omega_{pp}$ occurs at $k \lambda_D \approx 1$, $\omega = 0.76 \omega_{pp}$, and $\theta = 8^{\circ}$. 

Figures \ref{Disprelfig}c and \ref{Disprelfig}d show $\omega$ and $\gamma$ versus $k$ and $\theta$ for $V_\alpha = 200$~km~s$^{-1}$ and otherwise nominal conditions. Here, $\omega$ remains similar to the case for $V_\alpha = 100$~km~s$^{-1}$. However, $\gamma > 0$ occurs for $30^{\circ} < \theta < 73^{\circ}$, with $\gamma_{\mathrm{max}} = 0.12 \omega_{pp}$ occurring at $k \lambda_D \approx 1$ and $\theta = 60^{\circ}$. For these conditions, the ion-acoustic speed is $c_s \approx 100$~km~s$^{-1}$ and the phase speed $v_{ph}$ corresponding to $\gamma_{\mathrm{max}}$ is $v_{ph,\mathrm{max}}/c_s = 0.64$, indicating the waves are ion-acoustic-like waves. For $V_\alpha = 200$~km~s$^{-1}$, $v_{ph}$ is substantially smaller than $V_\alpha$, so waves aligned with ${\bf V}_{\alpha}$ are not growing, as they are not in Landau resonance with the alphas. Waves with oblique ${\bf k}$ become resonant by reducing the component of ${\bf k}$ aligned with ${\bf V}_{\alpha}$, such that $\omega/(k \cos{\theta}) \sim V_{\alpha}$. For $V_{\alpha} = 100$~km~s$^{-1}$ Landau resonance is maintained for $\theta = 0^{\circ}$, so waves oblique to ${\bf V}_\alpha$ are predicted for $V_\alpha \gtrsim v_{ph}$.

Figures \ref{Disprelfig}e--\ref{Disprelfig}g show $\gamma_{\mathrm{max}}$, $\omega_{\mathrm{max}}$, $k_{\mathrm{max}}$ versus $T_p$ and $T_\alpha$ for otherwise nominal conditions. For $V_\alpha = 100$~km~s$^{-1}$, $\gamma_{\mathrm{max}}$ peaks close to $\theta = 0^{\circ}$ so we only consider this case. We find that $\gamma_{\mathrm{max}}$ increases as $T_p$ and $T_\alpha$ decrease, although $\gamma > 0$ can occur for $T_p$ or $T_\alpha$ exceeding $T_e$ as long as the other ion component has temperature well below $T_e$. Fast growth, $\gamma \gtrsim 0.1 \omega_{pp}$, requires $T_p, T_\alpha \ll T_e$. Figures \ref{Disprelfig}f and \ref{Disprelfig}g show that $0.5 \lesssim \omega_{\mathrm{max}}/\omega_{pp} \lesssim 1$ and $k_{\mathrm{max}}\lambda_D \sim 1$ for most $T_p$ and $T_\alpha$ when instability occurs, so dispersion properties remain comparable. Figure \ref{Disprelfig}h shows $\gamma_{\mathrm{max}}$, $\omega_{\mathrm{max}}$, and $k_{\mathrm{max}}$ as a function of $n_\alpha/n_p$ and nominal conditions. Here, $n_\alpha/n_p$ is varied such that $n_p + 2 n_\alpha = n_e$ is satisfied and $n_e = 12$~cm$^{-3}$. We find that $ 0.8 \lesssim k_{\mathrm{max}} \lambda_D \lesssim 1$ and $\omega_{\mathrm{max}}/\omega_{pp} \gtrsim 0.6$. 
Peak growth $\gamma_{\mathrm{max}}$ increases with $n_\alpha/n_p$ until $n_\alpha/n_p \sim 1$. For the solar wind at 1~AU, $0.01 \lesssim n_\alpha/n_p \lesssim 0.1$, with average $n_\alpha/n_p \approx 0.04$ \cite{ogilvie1969,elliott2018}. Therefore, larger $n_\alpha/n_p$ should yield larger $\gamma$.

Figures \ref{Disprelfig}i--\ref{Disprelfig}l show $\gamma_{\mathrm{max}}$, $\omega_{\mathrm{max}}$, $\theta_{\mathrm{max}}$, and $k_{\mathrm{max}}$ as functions of $T_e$ and $V_{\alpha}$. The instability develops for $V_{\alpha} \gtrsim 60$~km~s$^{-1}$ and $T_e \gtrsim 20$~eV, with $\gamma$ increasing with $T_e$. Figure \ref{Disprelfig}j shows that $0.5 \lesssim \omega_{\mathrm{max}}/\omega_{pp} \lesssim 1$ for most of the parameter range where $\gamma > 0$. Figure \ref{Disprelfig}k shows that for low $V_\alpha$, $\theta_{\mathrm{max}} = 0^{\circ}$, but as $V_\alpha$ increases $\theta_{\mathrm{max}}$ increases to maintain Landau resonance with the alphas reaching $\theta_{\mathrm{max}} \sim 60^{\circ}$ for $V_\alpha \gtrsim 200$~km~s$^{-1}$. For a given $T_e$, $\gamma_{\mathrm{max}}$ increases with $V_\alpha$ until $\theta > 0^{\circ}$, whereupon $\gamma_{\mathrm{max}}$ remains constant as $V_\alpha$ and $\theta_{\mathrm{max}}$ increase. Figure \ref{Disprelfig}l shows that $k_{\mathrm{max}} \lambda_D \sim 1$, with $k_{\mathrm{max}} \lambda_D$ increasing slightly as $T_e$ increases for constant $V_\alpha$. 

In brief, ion-acoustic waves are generated by the proton-alpha streaming instability for a wide range of parameters relevant to the observations in section \ref{observationssec}.
Instability is found for $V_{\alpha} \gtrsim c_s/2$, either $T_p$ or $T_\alpha$ being significantly lower than $T_e$, and $n_\alpha/n_p$ typical of the solar wind. For $V_\alpha \sim v_{ph}$, the maximum $\gamma$ occurs for $\theta = 0^{\circ}$. For $V_\alpha \gtrsim v_{ph}$, $\theta_{\mathrm{max}}$ increases with $V_\alpha$, although the peak $\gamma$ is unchanged. For the shocks in section \ref{observationssec}, ion-acoustic waves should have ${\bf k}$ aligned with $\hat{\bf n}$, but as $V_\alpha = \Delta V_n$ increases, the waves become oblique to $\hat{\bf n}$. Since $T_e$ exceeding $T_p$ or $T_\alpha$ is required, the instability is favored close to the ramp, because downstream of the ramp $T_p$ or $T_\alpha$ may increase significantly. 



\section{Kinetic Simulation of the Proton-Alpha Instability} \label{secpicmodel}
We model the nonlinear evolution of the waves using a 1D electrostatic particle-in-cell (PIC) simulation  \cite{birdsall1991}, with the nominal plasma conditions from section \ref{seclintheory} and physical particle masses. The simulation has 128 grid points separated by $\Delta n = 0.5 \lambda_D$ and timestep of $\Delta t \omega_{pe} = 0.05$. The particle distributions are initialized as Maxwellians. Periodic boundary conditions are employed. We use $2 \times 10^4$ particles per cell per species. The simulation is run until $\omega_{pp} t = 255$, equivalent to $t = 0.061$~s. 

\begin{figure*}[htbp!]
\begin{center}
\includegraphics[width=160mm, height=142mm]{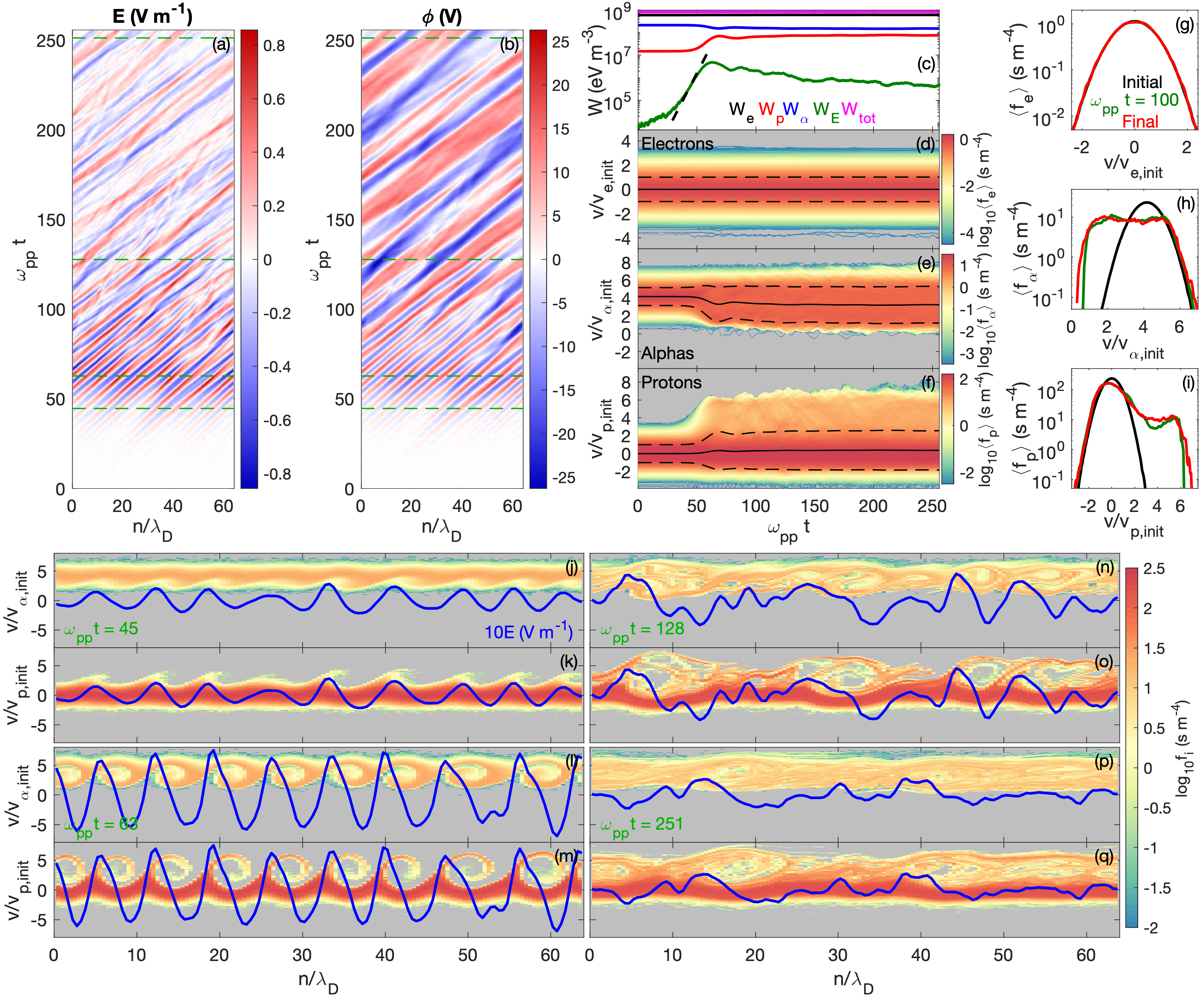}
\caption{PIC simulation of the proton-alpha streaming instability. (a) and (b) ${\bf E}$ and $\phi$ versus $n$ and $t$. (c) Spatially-averaged energy densities versus time. The black, red, blue, green, and magenta lines are the electron energy density $W_e$, proton energy density $W_p$, alpha energy density $W_\alpha$, ${\bf E}$ energy density $W_E$, and the total energy density $W_{tot}$, respectively. The black-dashed line is the linear theory prediction. (d)--(f) Spatially-averaged $\langle f_e \rangle$, $\langle f_\alpha \rangle$, and $\langle f_p \rangle$ versus time, respectively. The solid lines are $V_s$ and the dashed lines are $V_s \pm v_s$ for each species $s$. (g)--(i) $\langle f_e \rangle$, $\langle f_\alpha \rangle$, and $\langle f_p \rangle$ at the beginning and end of the simulation (black and red lines), and at $\omega_{pp} t = 100$ (green). (j)--(q) $f_\alpha$ and $f_p$ at four times indicated by the green dashed lines in panels (a) and (b). The blue lines indicate the corresponding $10 {\bf E}$. }
\label{PICsimfig}
\end{center}
\end{figure*}

The results are shown in Figure~\ref{PICsimfig}. Figures~\ref{PICsimfig}a--\ref{PICsimfig}b show ${\bf E}$ and the electrostatic potential $\phi$ versus $n$ and $t$. Periodic waves grow from noise, seen starting at $\omega_{pp} t \sim 30$. Figures \ref{PICsimfig}j and \ref{PICsimfig}k show the perturbations in the alpha and proton distributions, $f_p(n)$ and $f_\alpha(n)$, along $n$ at $\omega_{pp} t = 45$. The waves grow in time and reach saturation at $\omega_{pp} t \approx 63$ and are characterized by $k \lambda_D \approx 1$, consistent with Figure~\ref{Disprelfig}b. At saturation, ${\bf E}$ peaks at $\approx 800$~mV~m$^{-1}$. A quasi-periodic series of ion holes forms (Figures~\ref{PICsimfig}l--\ref{PICsimfig}m). Both protons and alphas become trapped in the negative potential wells. 

After saturation, the ion holes coalesce to form more isolated structures. Figures~\ref{PICsimfig}n--\ref{PICsimfig}o show the ion distributions at $\omega_{pp} t = 128$. Ion holes are observed around $n = 10 \lambda_D$ and $n = 30 \lambda_D$, with converging ${\bf E}$ and negative $\phi$. The ion distributions are complex due to prior coalescences. As ion holes coalesce, their widths increase, resulting in larger peak $\phi$ occurring after saturation. At the end of the simulation, ion holes in the proton and alpha distributions remain (Figures~\ref{PICsimfig}p--\ref{PICsimfig}q).

Figure~\ref{PICsimfig}c shows the changes in the spatially-averaged energy densities. Initially, the increase in the energy density of ${\bf E}$, $W_E$, is slow because the instability develops from noise. During the growth phase, $W_E$ increases approximately exponentially and approaches $\gamma_{\mathrm{max}}$ predicted from linear theory when ${\bf E}$ is well above the noise level. As the instability grows, the energy density $W_p$ of protons increases, while the energy density $W_\alpha$ of alphas decreases. After saturation, $W_E$ gradually decreases, while $W_p$ and $W_\alpha$ remain relatively constant. The energy density $W_e$ of electrons remains relatively constant throughout the simulation. Overall, we observe an energy transfer from drifting alphas to the fields and protons.

Figures \ref{PICsimfig}d--\ref{PICsimfig}f show the evolution of the spatially-averaged particle distributions $\langle f_e \rangle$, $\langle f_p \rangle$, and $\langle f_\alpha \rangle$. During the growth phase, $\langle f_\alpha \rangle$ broadens, with most of the particles losing speed as the instability grows, resulting in $V_\alpha$ decreasing and $T_\alpha$ increasing. For protons, a small fraction is in resonance with the waves as they grow and subsequently become trapped in the wave potentials. These protons gain energy, such that $\langle f_p \rangle$ forms a shoulder centered around $v_{ph}$ of the waves. 

The major changes in the distributions occur during the growth phase. 
After saturation, protons and alphas become trapped and untrapped in $\phi$ as the ion holes evolve and coalesce, but the spatially-averaged $\langle f_p \rangle$ and $\langle f_\alpha \rangle$ do not substantially change with time. Figures \ref{PICsimfig}g--\ref{PICsimfig}h show the initial distributions, final distributions, and distributions at $\omega_{pp} t = 100$. After $\omega_{pp} t = 100$ there is little change in the distributions. At the end of the simulation, $\langle f_\alpha \rangle$ has formed a broad flat-top distribution, while $\langle f_p \rangle$ forms a plateau over the same range of speeds as the alphas, resulting in marginal stability. For electrons, a flat-top  distribution develops at low energies, due to trapping between ion holes. Throughout the simulation $e \phi < T_e$, so only the lowest energy electrons are significantly affected. 

From $\langle f_e \rangle$, $\langle f_\alpha \rangle$, and $\langle f_p \rangle$ we calculate the particle moments. For electrons $T_e$ increases from $100$~eV to $102$~eV, so electron heating is negligible. For protons $V_p$ increases to $8.8$~km~s$^{-1}$ and $T_p$ increases from $3$~eV to $14.5$~eV. For alpha particles $V_\alpha$ decreases from $100$~km~s$^{-1}$ to $78$~km~s$^{-1}$ and $T_\alpha$ increases from $12$~eV to $50$~eV. Thus, the instability primarily affects the alpha particles due to conservation of momentum and their lower mass density. The primary effects of the instability are to heat the ions and reduce $\Delta V_n$.

\section{Discussion} \label{secdiscussion}
We investigated the proton-alpha streaming instability across low Mach number shocks. 
Linear theory shows that the instability is favored for $T_e > T_p$ or $T_\alpha$. Thus, the instability is likely to occur close to the ramp before the ions have been completely compressed and heated. Similarly, the instability, once active, will quickly saturate. This accounts for the largest amplitude waves occurring across the ramp, with their amplitude decreasing downstream. 

Previous studies have argued that ion-acoustic waves can be generated by the proton-proton streaming instability, Buneman instability, and instabilities associated with pressure gradients \cite{akimoto1985,allan1974,buneman1959}. For the five shocks considered here, two showed no evidence of proton reflection, so the proton-proton streaming instability cannot account for the waves. The shock widths varied by over an order of magnitude, substantially changing the peak current and gradients. Therefore, while it is possible that these instabilities could be active at some of the shocks, these instabilities are unlikely to explain the ion-acoustic waves at all five shocks. In contrast, large $\Delta {\bf V}$ occurred at all five shocks. Additionally, the ion-acoustic waves were highly oblique to ${\bf B}$ with wave vectors often aligned with the shock normal direction, consistent with the theoretical predictions for the proton-alpha streaming instability. 

We conclude that the proton-alpha streaming instability is likely occurring at the observed shocks and can account for the properties of the ion-acoustic waves.
Future work should include: 
\begin{itemize}
\item Investigating the proton-alpha streaming instability  for a wider range of shock parameters, in particular, supercritical shocks typical of Earth's bow shock. 
\item In section \ref{seclintheory} we assumed the unmagnetized dispersion equation. Magnetized electrons should be considered in detail and how this modifies the instability properties. 
\item For large $\Delta {\bf V}$, wave growth is oblique, which requires 2D or 3D simulations to accurately model. This may modify the evolution of the ion distributions. 
\end{itemize}

\section{Conclusions} \label{secconclusions}
In this letter, we investigated the development of ion-acoustic waves at low Mach number shocks. The key results are: 
\begin{itemize}
    \item The cross-shock potential results in protons being decelerated more than alpha particles, producing a large drift between the ion distributions across the ramp. This relative drift can be unstable to the proton-alpha streaming instability, producing large-amplitude ion-acoustic waves. 
    \item Linear theory shows that the ion-acoustic waves can be generated for a wide range of plasma parameters, although electron temperature exceeding ion temperatures is required. The predicted and observed wave properties are in good agreement. 
    \item The effect of the instability is to reduce the relative drift between proton and alpha distributions and heat both ion species. Protons are predicted to form a shoulder in the direction of the relative alpha velocity, while alpha particles form a flat-top distribution. 
\end{itemize}
We propose that the proton-alpha streaming instability is the likely source of ion-acoustic waves at the bow shock crossings observed on 24 April 2023. The relative drift between protons and alphas produced by the cross-shock potential should be a characteristic feature for a wide range of shock parameters, and thus the proton-alpha streaming instability could drive ion-acoustic waves at high Mach number shocks. 

\section*{Data Availability Statement}
MMS data are available at \url{https://lasp.colorado.edu/mms/sdc/public} and \url{https://spdf.gsfc.nasa.gov/pub/data/mms/}. We use burst mode magnetic field data from FGM \cite{mms1fgmbrst}, electric field data \cite{mms1edpdcebrst} and probe potential data from EDP \cite{mms1edpscpotbrst}. We use burst mode ion distributions from FPI \cite{mms1fpidisdistbrst} and burst mode electron moments from FPI \cite{mms1fpidesmomsbrst}. The data analysis was performed using the irfu-matlab software package \cite{khotyaintsev_2024_11550091}. The scripts required to generate the model data and figures in this paper are available at \url{https://doi.org/10.5281/zenodo.14828056} \cite{graham2025}.

\acknowledgments
We thank the MMS team and instrument PIs for data access and support. This work was supported by the Swedish National Space Agency (SNSA), Grant 206/19. 


\end{document}